\documentclass[conference]{IEEEtran}
\IEEEoverridecommandlockouts
\usepackage{cite}
\usepackage{amsmath,amssymb,amsfonts}
\usepackage{algorithmic}
\usepackage{graphicx}
\usepackage{tabularx}
\usepackage{comment}
\usepackage{booktabs}
\usepackage{threeparttable}
\usepackage{multirow}
\usepackage{siunitx}
\usepackage{relsize}
\usepackage[caption=false,font=normalsize,labelfont=sf,textfont=sf]{subfig}
\usepackage{textcomp}
\usepackage{xcolor}
\def\BibTeX{{\rm B\kern-.05em{\sc i\kern-.025em b}\kern-.08em
    T\kern-.1667em\lower.7ex\hbox{E}\kern-.125emX}}

\renewcommand{\baselinestretch}{0.9831}
  
\begin{document}

\title{
AC Power Flow Informed Parameter Learning\\ for DC Power Flow Network Equivalents\\

\thanks{ \IEEEauthorrefmark{1a} School of Electrical and Computer Engineering, Georgia Institute of Technology. \{taheri, molzahn\}@gatech.edu. Support from NSF award \#2145564.}
}

\author{\IEEEauthorblockN{Babak Taheri\IEEEauthorrefmark{1} and Daniel K. Molzahn\IEEEauthorrefmark{1}}
}

\maketitle

\begin{abstract}

This paper presents an algorithm to optimize the parameters of power systems equivalents to enhance the accuracy of the DC power flow approximation in reduced networks. Based on a zonal division of the network, the algorithm produces a reduced power system equivalent that captures inter-zonal flows with aggregated buses and equivalent transmission lines. The algorithm refines coefficient and bias parameters for the DC power flow model of the reduced network, aiming to minimize discrepancies between inter-zonal flows in DC and AC power flow results. Using optimization methods like BFGS, L-BFGS, and TNC in an offline training phase, these parameters boost the accuracy of online DC power flow computations. In contrast to existing network equivalencing methods, the proposed algorithm optimizes accuracy over a specified range of operation as opposed to only considering a single nominal point. Numerical tests demonstrate substantial accuracy improvements over traditional equivalencing and approximation methods.
\end{abstract}


\section{Introduction}
The complexity of modern power systems significantly challenges system planning and analysis, necessitating computational resources that are often prohibitive in scale. While using AC power flow equations with fully detailed network models provides high accuracy \cite{cheng2005ptdf, Wang2007}, the sheer size of many systems renders this approach computationally burdensome for many applications.
These computational challenges are often addressed via two strategies: employing DC power flow approximations and adopting reduced equivalents. The DC power flow model's linearity enables the application of linear programming solvers, often making this model a practical choice despite its simplified representation of network behavior.
Moreover, network equivalencing is a critical tool across various power system applications, such as market analysis and planning problems. By condensing large, detailed power networks into smaller, more manageable models, equivalencing aims to simplify analyses without substantial loss of accuracy.

The literature includes many network equivalencing methods with different advantages and disadvantages~\cite{ward1949equivalent, dimo1975nodal, allen2008combined, oh2009new, cheng2005ptdf, shi2012improved, shi2014novel}; see \cite{Gupta2018} for a review. A common limitation of these methods is their focus on a specific operating point. This approach is appropriate for some applications which have good estimates of nominal operations but is less useful for others where accuracy across a broader operating range is critical, such as optimization problems where the base case operating point is not predetermined. Furthermore, these methods attempt to match the \emph{DC power flow} model's solution in the original network, rather than reflecting the actual power flows of the lines as modeled in \emph{AC power flow} solutions. Our algorithm addresses these gaps by ensuring accuracy over a wider range of operating conditions, offering significant advantages in applications including planning and market analyses.

Methods similar to Ward reduction, for instance, disperse generator injections among several boundary buses, which is not practical for certain simulation tools \cite{shi2014novel}. A modification presented in \cite{allen2008combined} attempts to preserve full generator models at the retained buses, but this may still misrepresent certain power flows, leading to erroneous interchange analyses.
Similarly, the REI equivalent method \cite{dimo1975nodal} does not maintain model sparsity during substantial size reductions, leading to a densely connected bus susceptance matrix and introducing fictitious branches between zones, making it inappropriate for some system planning and power market analyses.

The technique presented in \cite{cheng2005ptdf} endeavors to approximate the system's Power Transfer Distribution Factors (PTDFs), which relate power injections at buses to power flows on lines. However, this technique may not accurately align with zonal PTDFs, especially under varying operating points. The method in \cite{oh2009new} aims to replicate the original system's inter-zonal power flows, primarily at the base case.
Similarly, the approach in \cite{shi2014novel} treats network reduction as a quadratic optimization problem, striving to match inter-zonal flows with those of the original system's DC power flow model.
By focusing on the base case, existing methods do not consider a broader range of operation. Typically, these methods aim to match the inter-zonal DC power flows of the original system. None of the existing methods address our objective of matching inter-zonal AC power flows in the original network over a range of operations.

In this paper, we propose a network equivalencing algorithm designed to match the solution of DC power flow in the reduced network with the \emph{AC power flow} solution of the original network over a range of operations. This approach leverages our prior work in optimally selecting the parameters in DC power flow approximations based on their performance over a range of operational conditions, as detailed in \cite{taheri2023optimizing}. Building upon these techniques, we extend our algorithm to address the equivalencing problem considered in this paper.
Our algorithm optimizes coefficient and bias parameters in the reduced network, aiming to align its solution more closely with the inter-zonal flows in the AC power flow model for the original network. Our approach takes as input a segmentation of the original network into specified zones. 
The power system equivalents aggregate buses in each zone and form equivalent transmission lines between zones.
Using gradient-based optimization techniques, our algorithm has an offline training phase to compute optimal network parameters. These optimized parameters are subsequently deployed to bolster the accuracy of inter-zonal flows derived from DC power flow estimations in online computations. This parameter optimization results in solutions from the reduced network more closely matching the AC power flow solutions of the original network.

In summary, the main contributions of this paper are:
\begin{enumerate}
    \item Based on specified zones, we compute parameters for the reduced network to more closely match the \emph{AC power flow} model for the original network across a range of power injection scenarios.
    \item We use various numerical methods such as BFGS, \mbox{L-BFGS}, and TNC in a stochastic fashion to scale our proposed algorithm to large power systems.
    \item We showcase numerical results that underscore the superior accuracy of our proposed algorithm.
\end{enumerate}

The rest of the paper is organized as follows: Section~\ref{sec:Network Reduction} formulates our network reduction problem. Section~\ref{sec:Optimizing Parameters} discusses the algorithm we propose for optimizing the parameters of the reduced network. Section~\ref{sec:Numerical Results} shows and compares the results of the proposed algorithm. Section~\ref{sec:Conclusion} concludes the paper.

\section{Network Reduction}
\label{sec:Network Reduction}

\begin{figure}[htbp]
\centerline{\includegraphics[width=0.485\textwidth]{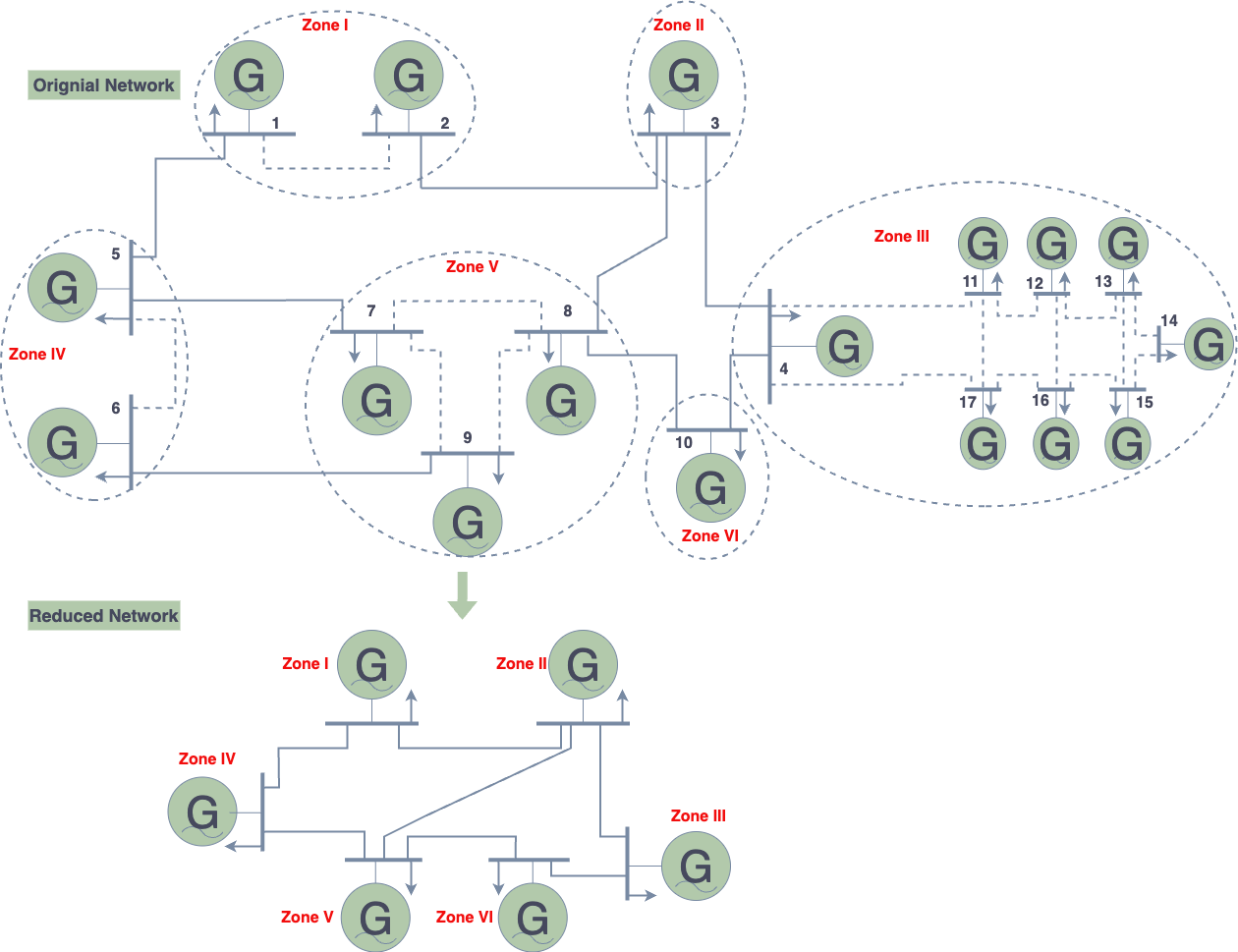}}
\vspace{-0.5em}
\caption{Topology of a 17-bus network with six zones and its reduced equivalent. In this depiction, zones are reduced to single buses with aggregated generators, representing total zone capacity. Note that in specific applications, each bus in the reduced network may connect to multiple original generators to preserve individual characteristics.}
\label{fig:reduction}
\vspace{-1em}
\end{figure}

We aim to compute a reduced network with a DC power flow approximation that matches the inter-zonal flows of the original network with the AC power flow equations. In our algorithm, we start with pre-defined zones within the original network, which could be based on various factors such as congestion areas or utility service territories.
Pairs of zones in the reduced network are interconnected with a transmission line only if they were connected in the original network. Refer to Fig.~\ref{fig:reduction} for an illustrative example where the original network is segmented into six zones, each encircled by a dashed line. Each zone is represented by a single bus in the reduced network, with the aggregated generation and loads attached to the corresponding equivalent buses.\footnote{In the network reduction process, generators within each zone can be handled in two ways: either by attaching all generators to their corresponding bus in the reduced network or by aggregating them into a single generator with equivalent output capabilities. This choice does not impact our network reduction process.}

We next define our notation. Use $\mathcal{N}$, $ref$, and $\mathcal{E}$ for sets of buses, the reference bus, and lines in the original network, respectively. Each bus $i\in\mathcal{N}$ has a voltage phasor with magnitude $V_i$ and phase angle $\theta_{i}$ as well as a complex power injection $S_i = P_i + jQ_i$ and a shunt admittance $Y_i^S$. Active and reactive power flows on the line $(i,j)\in\mathcal{E}$ are denoted by $p_{ij}$ and $q_{ij}$, respectively.
Each line $(i,j)\in\mathcal{E}$ has series admittance $Y_{ij}$ and shunt admittance $Y_{ij}^{sh}$. Real and imaginary parts of complex values are given by $\Re(\,\cdot\,)$ and $\Im(\,\cdot\,)$ respectively. The matrix transpose is $(\,\cdot\,)^T$.
For the reduced network, $\mathcal{N}_R$, $\mathcal{E}_R$, and $ref_R$ denote the sets of buses (equivalent to the zones in the original network), lines, and the reference bus.

\subsection{AC Power Flow}
The AC power flow equations model a power system through non-linear relationships among voltage magnitudes, phase angles, and complex power flows and injections.
The AC power flow equations are:
\begin{subequations}\label{eq:AC_PF}%
\begin{align}%
\raisetag{0.75em} P_i =& \sum_{(i,j) \in \mathcal{E}} p_{ij} + V_i^2 \Re(Y^{S}_i), ~~~
Q_i = \sum_{(i,j) \in \mathcal{E}} q_{ij} - V_i^2 \Im(Y^{S}_i), \\
p_{i j} =& V_i^2\left(\Re(Y_{ij})+\Re(Y_{ij}^{sh})\right)- V_i V_j \Re(Y_{ij}) \cos (\theta_{i}-\theta_{j})\nonumber
&\\ 
& \quad  - V_i V_j \Im(Y_{ij}) \sin (\theta_{i}-\theta_{j}), \\
q_{i j} =& -V_i^2\left(\Im(Y_{ij}) +\Im(Y_{ij}^{sh}) \right) - V_i V_j \Re(Y_{ij})  \sin (\theta_{i}-\theta_{j})\nonumber &\\
& \quad + V_i V_j\Im(Y_{ij})  \cos (\theta_{i}-\theta_{j}).
\end{align}
\end{subequations}

\subsection{Inter-Zonal Flows}
After solving \eqref{eq:AC_PF}, we can calculate the inter-zonal flows from the AC power flow solution for the original network. The term inter-zonal power flow denotes the aggregate power transfer between two distinct zones. In the original network, inter-zonal power flows are computed by adding all the power flows on all transmission lines linking the respective zones:
\begin{equation}
\label{eq:}
    \mathbf{p}^{iner-zonal} = \mathbf{\Psi}_{flow} \cdot \mathbf{p}^{AC}
\end{equation}
where $\mathbf{\Psi}_{flow}$ is a $|\mathcal{E}_R| \times |\mathcal{E}|$ matrix that sums up the line flows, $\mathbf{p}^{iner-zonal}$ is a length-$|\mathcal{E}_R|$ vector of the inter-zonal flows, and $\mathbf{p}^{AC}$ is a length-$|\mathcal{E}|$ vector of the line flows from the AC power flow solution.

\subsection{DC Power Flow}
The DC power flow approximation simplifies the AC equations by neglecting reactive power, assuming voltage magnitudes are fixed to a nominal value, and assuming only minor angle differences across lines to apply the small angle approximation. The DC flow results from applying these assumptions to the AC power flow equations~\eqref{eq:AC_PF}:
\begin{subequations}
\label{eq:DCPF}
\begin{align}
P_i - \gamma_{i}&= \sum_{j \in \mathcal{N}} b_{ij} \cdot (\theta_i - \theta_j),
\label{eq:PowerBalance}\\
p^{DC}_{ij} &= b_{ij} \cdot (\theta_{i}-\theta_{j}) + \rho_{ij},
\label{eq:Line_flow}
\end{align}
\end{subequations}
where $p^{DC}_{ij}$ denotes the power flow in line $(i,j) \in \mathcal{E}$. As expounded in~\cite{Stott2009}, $\gamma_i$ is a bias parameter that compensates for losses originating from shunts, HVDC infeeds, and injections which model phase shifts and branch losses for lines connected to bus~$i$. The bias parameter $\rho_{ij}$ for line $(i,j) \in \mathcal{E}$ is associated with line losses.

Let $\mathcal{N}^{\prime} = \mathcal{N} \setminus {ref}$ denote the set of all buses apart from the reference bus. Let $\mathbf{P}$ and $\boldsymbol{\theta}$ denote the vectors of power injections and voltage angles at each bus $i \in \mathcal{N}^{\prime}$. Set $\theta_{ref} = 0$. In addition, define $\mathbf{A}$ as the $|\mathcal{E}| \times (|\mathcal{N}|-1)$ node-arc incidence matrix detailing the linkages between the system's buses and branches, and let $\mathbf{b}$ be a length-$|\mathcal{E}|$ coefficient vector commonly derived using the branch susceptances.
The matrix form of~\eqref{eq:DCPF} is:
\begin{subequations}
\label{eq:DCPF_matrix}
\begin{align}
\mathbf{P} - \boldsymbol{\gamma}&= \mathbf{B}^{\prime} \cdot \boldsymbol{\theta},
\label{eq:MatrixForm}\\
\mathbf{p}^{DC} &= \left(\text{diag}(\mathbf{b}) \cdot \mathbf{A} \cdot \boldsymbol{\theta}\right) +\boldsymbol{\rho},
\label{eq:branchPowerFlow}
\intertext{where $\text{diag}(\,\cdot\,)$ is the diagonal matrix with the argument on the diagonal and $\mathbf{B}^{\prime}$ is}
\mathbf{B}^{\prime} &= \mathbf{A}^T \cdot \text{diag}(\mathbf{b}) \cdot \mathbf{A}. \label{eq:B-prime_sub}
\end{align}
\end{subequations}
In \eqref{eq:DCPF_matrix}, $\mathbf{p}^{DC}$ represents a length-$|\mathcal{E}|$ vector of power flows across each branch and $\boldsymbol{\rho}$ is a length-$|\mathcal{E}|$ vector associated with line losses.

By solving~\eqref{eq:MatrixForm} for $\boldsymbol{\theta}$ and substituting into~\eqref{eq:branchPowerFlow}, the resulting expression is the PTDF formulation of the DC power flow equations, which linearly relates the line flows to the active power injections:
\begin{equation}
\mathbf{p}^{DC} = \text{diag}(\mathbf{b}) \cdot \mathbf{A} \cdot [\mathbf{A}^T \cdot \text{diag}(\mathbf{b}) \cdot \mathbf{A}]^{-1}\cdot (\mathbf{P} - \boldsymbol{\gamma}) + \boldsymbol{\rho}.
\label{eq:branchPowerFlow2}
\end{equation}
The parameters $\mathbf{b}$, $\boldsymbol{\gamma}$, and $\boldsymbol{\rho}$ strongly influence the DC power flow's accuracy.

\subsection{DC Power Flow in the Reduced Network}

We can use the DC power flow approximation as shown in \eqref{eq:branchPowerFlow2} in the reduced network by replacing the set of buses ($\mathcal{N}$) with $\mathcal{N}_R$ (number of zones), set of lines ($\mathcal{E}$) with $\mathcal{E}_R$, and reference bus with reference zone (the zone that contains the reference bus). Furthermore, we replace the $\mathbf{A}$ matrix with $\mathbf{A}_R$ as the $|\mathcal{E}_R| \times (|\mathcal{N}_R|-1)$ node-arc incidence matrix for the reduced network, and $\mathbf{P}$ with $\mathbf{P}_R$ as a length-($|\mathcal{N}_R|-1$) vector representing the power injections at each zone excluding the reference zone. We also replace the $\mathbf{b}$, $\boldsymbol{\gamma}$, and $\boldsymbol{\rho}$ parameters with $\mathbf{b}_R$, $\boldsymbol{\gamma}_R$, and $\boldsymbol{\rho}_R$ as length-$|\mathcal{E}_R|$, length-($|\mathcal{N}_R-1|$), and length-$|\mathcal{E}_R|$ vectors, respectively.
Finally, power injections at each bus/zone in the reduced network (i.e., $\mathbf{P}_R$ is a length-($|\mathcal{N}_R|-1$) vector) equals the aggregated power injections at the related buses within each zone in the initial system, which can be calculated as follows:
\begin{equation}
 \mathbf{P}_{R} = \mathbf{\Psi}_{g} \times \mathbf{P},
\end{equation}
where $\mathbf{\Psi}_{g}$ is a $|\mathcal{N}_R| \times |\mathcal{N}|$ zone--bus connection matrix (i.e., $\Psi^{i,j}_{g}$ is equal to $1$ if bus $j$ belongs to zone $i$). Accordingly, \eqref{eq:branchPowerFlow2} can be revised as follows for the reduced network:
\begin{equation}
\mathbf{p}^{DC}_{R} = \text{diag}(\mathbf{b}_{R}) \cdot \mathbf{A}_{R} \cdot [\mathbf{A}_{R}^T \cdot \text{diag}(\mathbf{b}_{R}) \cdot \mathbf{A}_{R}]^{-1}\cdot (\mathbf{P}_{R} - \boldsymbol{\gamma}_{R}) + \boldsymbol{\rho}_{R}.
\label{eq:branchPowerFlow_reduced}
\end{equation}
Next, we present a machine learning inspired algorithm to refine the coefficient ($\mathbf{b}_R$) and bias ($\boldsymbol{\gamma}_R$ and $\boldsymbol{\rho}_R$) parameters in the reduced network, targeting a closer match to inter-zonal flows from the AC solutions.

\begin{figure}[b]
\vspace{-1em}
\centerline{\includegraphics[width=0.485\textwidth]{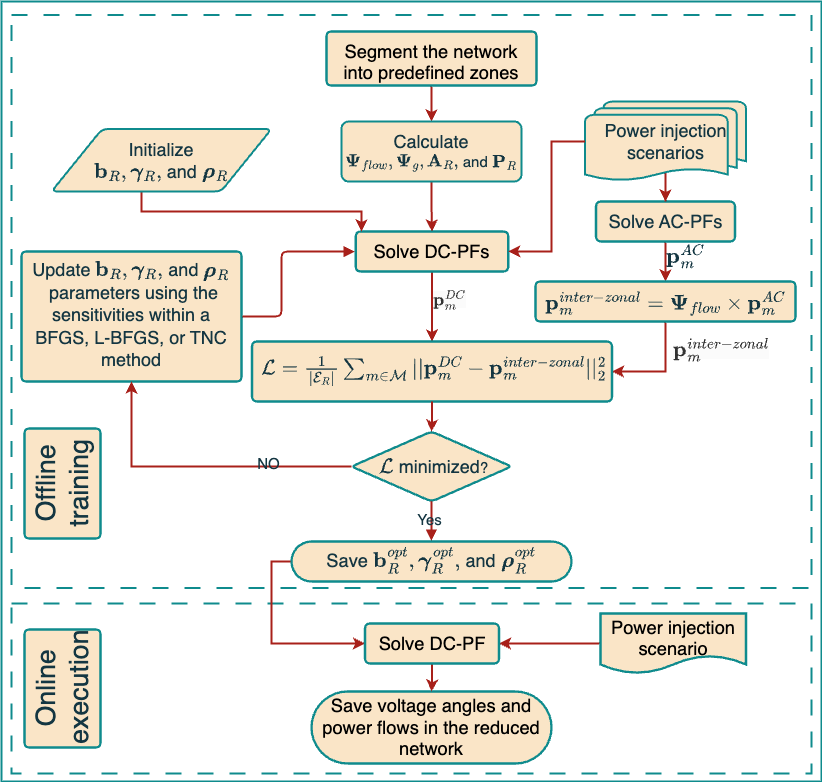}}
\vspace{-0.7em}
\caption{Flowchart of the proposed algorithm}
\label{fig:flowchart}
\end{figure}

\section{Optimizing the Reduced\\ Network's Parameters}
\label{sec:Optimizing Parameters}
Fig.~\ref{fig:flowchart} illustrates our parameter optimization algorithm, divided into \textit{offline} and \textit{online} phases. Building upon the algorithm introduced in our previous work \cite{taheri2023optimizing}, this study extends its application to learning the parameters of equivalent networks.
The \textit{offline} phase, executed once, computes the optimal values for the $\mathbf{b}_R$, $\boldsymbol{\gamma}_R$, and $\boldsymbol{\rho}_R$ parameters across a set of power injection scenarios such that the inter-zonal flows in our reduced DC model align well with the inter-zonal AC power flows in the original system under a range of operational conditions. Conversely, the \textit{online} phase leverages these optimized parameters (i.e., $\mathbf{b}^{opt}_R$, $\boldsymbol{\gamma}^{opt}_R$, and $\boldsymbol{\rho}^{opt}_R$) in the reduced network to improve accuracy and computational speed during real-time operations. Hence, the algorithm allocates computational effort in the preliminary offline optimization to achieve accuracy advantages during online executions.
The primary advantage of this algorithm lies not merely in solving DC power flow problems but in facilitating the application of these optimized parameters in more complex power system problems such as optimal power flow and unit commitment.

Our algorithm is inspired by a supervised machine learning paradigm that uses power injections as inputs and inter-zonal line flows in the reduced network as targets. However, we emphasize that our algorithm does not use traditional machine learning models such as neural networks. Instead, the offline phase fine-tunes the parameters $\mathbf{b}_R$, $\boldsymbol{\gamma}_R$, and $\boldsymbol{\rho}_R$ through an optimization problem.
The optimization begins with a loss function that compares the inter-zonal flows from the DC power flow in the reduced network against the inter-zonal flows from the AC power flow equations in the original network across various power injection scenarios. Sensitivities of the loss function concerning these parameters are then computed.
Informed by these sensitivities, optimization methods like BFGS, L-BFGS, and TNC are employed, offering scalable optimization suitable for large power systems. Our algorithm fine-tunes the parameters, significantly enhancing the DC power flow model accuracy in the reduced network across diverse power injection scenarios.

\subsection{Loss Function}
\label{subsec:Loss Function}
We propose a loss function, $\mathcal{L}$, based on the sum of squared two-norm discrepancies between AC inter-zonal flows ($\mathbf{p}_{m}^{inter-zonal}$) and DC inter-zonal flows ($\mathbf{p}_{R,m}^{DC}$) across specified power injection scenarios $m \in \mathcal{M} = {1, 2, \ldots, S}$. This approach, common in machine learning for its robustness and differentiability, is expressed as:
\begin{align}
\nonumber \mathcal{L} &= \frac{1}{|\mathcal{E}_R|}\sum_{m \in \mathcal{M}} ||\mathbf{p}_{R,m}^{DC} - \mathbf{p}_{m}^{inter-zonal}||^2_2\\ %
& = \frac{1}{|\mathcal{E}_R|}\sum_{m \in \mathcal{M}} ||\text{diag}(\mathbf{b}_R) \cdot \mathbf{A}_R \cdot [\mathbf{A}_R^T \cdot \text{diag}(\mathbf{b}_R)\cdot\mathbf{A}_R]^{-1} \nonumber\\ \label{eq:objective_function}%
& \qquad \qquad \qquad \times  (\mathbf{P}_{R,m} - \boldsymbol{\gamma}_{R}) +\boldsymbol{\rho}_{R} - \mathbf{p}_{m}^{inter-zonal}||^2_2, 
\end{align}%
where the constant $\frac{1}{|\mathcal{E}_R|}$ normalizes the function by the system size. As indicated in \eqref{eq:branchPowerFlow_reduced} and \eqref{eq:objective_function}, the DC power flows $\mathbf{p}_{R,m}^{DC}$, and thus $\mathcal{L}(\mathbf{b}_R, \boldsymbol{\gamma}_R,\boldsymbol{\rho}_R)$, depend on parameters $\mathbf{b}_R$, $\boldsymbol{\gamma}_R$, and $\boldsymbol{\rho}_R$. By emphasizing larger deviations, the squared two-norm formulation aligns with applications where severe errors are especially undesirable. Alternative norms could be utilized with minimal conceptual alterations.

To find the optimized parameters, the algorithm solves:
\begin{equation}
\min_{\mathbf{b}_R,\boldsymbol{\gamma}_R,\boldsymbol{\rho}_R} \quad \mathcal{L}(\mathbf{b}_R,\boldsymbol{\gamma}_R,\boldsymbol{\rho}_R).
\label{eq:optimization problem}
\end{equation}

\subsection{Analysis of Coefficient and Bias Parameters' Sensitivities}
\label{subsec:Sensitivity Analysis}
To solve~\eqref{eq:optimization problem}, we use optimization techniques such as BFGS, L-BFGS, and TNC that depend on the gradient of the loss function relative to the parameters \( \mathbf{b}_R, \boldsymbol{\gamma}_R, \) and \( \boldsymbol{\rho}_R \). We next analyze the sensitivity of the loss function \( \mathcal{L}(\mathbf{b}_R,\boldsymbol{\gamma}_R, \boldsymbol{\rho}_R) \) to changes in these parameters across different power injection scenarios. We first focus on the sensitivities to the \( \mathbf{b}_R \) parameters, symbolized as \( \mathbf{g}^{b_R} \), derived by differentiating the loss function with respect to the coefficient parameters \( \mathbf{b} \):
\begin{subequations}
\label{eq:sensitivity}
\begin{equation}
\label{eq:sensitivity1}
 \mathbf{g}^{b_R}= \frac{2}{|\mathcal{E}|}\sum_{m \in \mathcal{M}} \left. \frac{\partial \mathbf{p}_{R}^{DC}}{\partial \mathbf{b}_R} \right|_ {\mathbf{p}^{DC}_{R,m}}\Big(\mathbf{p}_{R,m}^{DC} - \mathbf{p}_{m}^{inter-zonal} \Big),
\end{equation}
where $\frac{\partial \mathbf{p}_{R}^{DC}}{\partial \mathbf{b}_R}$ is obtained from the derivative of \eqref{eq:branchPowerFlow_reduced} with respect to the coefficient parameters $\mathbf{b}_R$,
\begin{align}
\frac{\partial \mathbf{p}_{R}^{DC}}{\partial \mathbf{b}_R} = & \text{~diag} \Bigg(\Big [\mathbf{A}_R [\mathbf{A}_R^T \cdot \text{diag}(\mathbf{b}_R) \cdot \mathbf{A}_R]^{-1} (\mathbf{P}_R-\boldsymbol{\gamma}_R) \Big]^{T} \Bigg) \times \nonumber \\
&  \Big (\mathbf{I} - \text{diag}(\mathbf{b}_R) \cdot \mathbf{A}_R [\mathbf{A}_R^T \cdot \text{diag}(\mathbf{b}_R) \cdot \mathbf{A}_R]^{-1} \mathbf{A}_R^T \Big),
\label{eq:coef_sensitivity}
\end{align}
\end{subequations}
and $\mathbf{I}$ is the identity matrix. The appendix in \cite{taheri2023optimizing} provides a detailed derivation of~\eqref{eq:coef_sensitivity} for the original network.

Similarly, the gradient of the loss function with respect to the \( \boldsymbol{\gamma}_R \) parameters is represented by \( \mathbf{g}^{\gamma_R} \):
\begin{subequations}
\begin{equation}
\label{eq:sensitivity-bias}
 \mathbf{g}^{\gamma_{R}}= \frac{2}{|\mathcal{E}|}\sum_{m \in \mathcal{M}} \left. \frac{\partial \mathbf{p}_{R}^{DC}}{\partial \boldsymbol{\gamma}_{R}} \right|_ {\mathbf{p}^{DC}_{R,m}}\Big(\mathbf{p}_{R,m}^{DC} - \mathbf{p}_{m}^{inter-zonal} \Big),
\end{equation}
where $\frac{\partial \mathbf{p}_{R}^{DC}}{\partial \boldsymbol{\gamma}_{R}}$ is calculated by taking the derivative of \eqref{eq:branchPowerFlow_reduced} with respect to bias parameters $\boldsymbol{\gamma}_{R}$:
\begin{align}
\frac{\partial \mathbf{p}_{R}^{DC}}{\partial \boldsymbol{\gamma}_{R}} = -\text{diag}(\mathbf{b}_{R}) \cdot \mathbf{A}_{R} [\mathbf{A}_{R}^T \cdot \text{diag}(\mathbf{b}_{R}) \cdot \mathbf{A}_{R}]^{-1}.
\label{eq:coef_sensitivity2}
\end{align}
\end{subequations}

Lastly, the sensitivities associated with the \( \boldsymbol{\rho}_R \) parameters are introduced, symbolized by \( \mathbf{g}^{\rho}_{R} \):
\begin{subequations}
\begin{equation}
\label{eq:sensitivity-rho}
 \mathbf{g}^{\rho_{R}}= \frac{2}{|\mathcal{E}|}\sum_{m \in \mathcal{M}} \left. \frac{\partial \mathbf{p}_{R}^{DC}}{\partial \boldsymbol{\rho}_{R}} \right|_ {\mathbf{p}^{DC}_{R,m}}\Big(\mathbf{p}_{R,m}^{DC} - \mathbf{p}_{m}^{inter-zonal} \Big),
\end{equation}
where \(\frac{\partial \mathbf{p}_{R}^{DC}}{\partial \boldsymbol{\rho}_{R}}\) is calculated by taking the derivative of \eqref{eq:branchPowerFlow_reduced} with respect to bias parameters \(\boldsymbol{\rho}_{R}\), which is the identity matrix \(\mathbf{I}\).
\end{subequations}
These sensitivities are used in gradient-based methods to compute the parameters \( \mathbf{b}_{R}, \boldsymbol{\gamma}_{R}, \) and \( \boldsymbol{\rho}_{R} \), as discussed next.

\subsection{Gradient-Based Optimization Methods}
\label{sec:Optimization Methods}

This subsection focuses on employing various gradient-based methods for tackling the optimization problem~\eqref{eq:optimization problem}, given the sensitivities in Section~\ref{subsec:Sensitivity Analysis}. Several methods like BFGS, L-BFGS, and TNC are discussed here and are later empirically evaluated using multiple test cases.

\begin{enumerate}
    \item \textbf{BFGS}: This iterative method, conceptualized by Broyden, Fletcher, Goldfarb, and Shanno~\cite[p.~136]{nocedal2006numerical}, utilizes gradient information to iteratively refine an inverse Hessian matrix approximation, eliminating the requirement for a full Hessian matrix.
    
    \item \textbf{L-BFGS}: As an advancement of the BFGS algorithm, the L-BFGS adopts a limited memory strategy to better manage large datasets.

    \item \textbf{Truncated Newton Conjugate-Gradient (TNC)}: The TNC method deploys a truncated Newton algorithm for minimizing a function with bound-constrained variables~\cite{nocedal2006numerical, nash1984newton}.
\end{enumerate}

Using the stochastic versions of these optimization methods, we can optimize the coefficient and bias parameters in the power system equivalents for large-scale systems.

\section{Experimental Results and Discussion}
\label{sec:Numerical Results}

This section showcases and assesses the results from our proposed algorithm. To demonstrate the model's effectiveness, we contrast the power flows from our machine learning-inspired algorithm against those from traditional network reduction algorithms. 
In these test cases, we generated $10,000$ power injection scenarios, allocating $8,000$ for offline training and $2,000$ for testing purposes. The batch size for the stochastic optimization methods is set to $100$. These scenarios were formulated by multiplying the nominal power injections by a normally distributed random variable with a mean of zero and a standard deviation of $15\%$.
The solutions to the AC power flow problems were computed utilizing \texttt{PowerModels.jl}\cite{coffrin2018powermodels} on a computing node within the Partnership for an Advanced Computing Environment (PACE) cluster at Georgia Tech. This node is equipped with a 24-core CPU and 32GB of RAM. Our algorithm is implemented in Python 3.0 within a Jupyter Notebook environment. For the minimization of the loss function, we employed the BFGS, \mbox{L-BFGS}, TNC, methods from the \texttt{scipy.optimize.minimize} library.

\subsection{Small 6-Bus Example}
\begin{figure}[b]
\centerline{\includegraphics[width=0.485\textwidth]{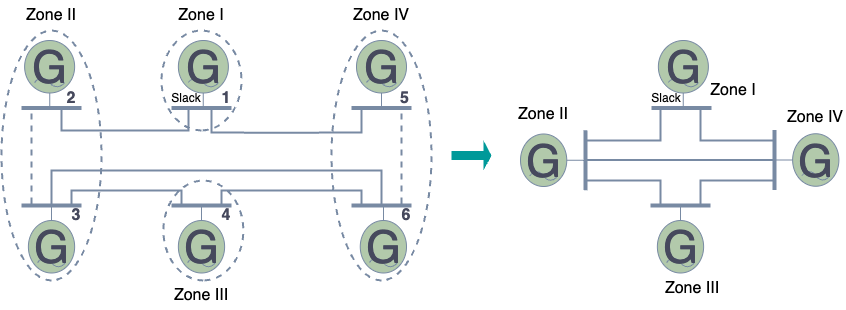}}
\vspace{-1em}
\caption{The original 6-bus network and the resulting 4-bus equivalent}
\label{fig:original 6-bus}
\end{figure}

Here, we show the performance of the proposed algorithm using a small 6-bus network as shown in Fig.~\ref{fig:original 6-bus} and compare it to the previously proposed algorithms. 
For simplicity, all the transmission line reactances in this network are designated as $j0.1$ per unit, matching the system used in~\cite{oh2009new, cheng2005ptdf, shi2012improved, shi2014novel}. The system's base power is $100$ MVA, and bus $1$ is chosen as the slack bus. Finally, the base case power injections are $-400, 100, 200, 50, 30, 20$~MW for buses $1$ to $6$, respectively. We generated $10,000$ power injection scenarios, with $8,000$ for training and $2,000$ for testing. Each scenario was created by multiplying nominal power injections with a normally distributed random variable, having a mean of zero and a standard deviation of $15\%$.
Fig.~\ref{fig:original 6-bus} shows the original network with four zones as well as the resulting four-bus equivalent.

We first apply the proposed algorithm using the training dataset, resulting in the optimized parameters given in Table~\ref{tab:parameters}.
Then, we solve the DC power flow model of the reduced network using the given injection scenario to compare it to the existing literature. 
Table~\ref{tab:comparison} illustrates the results of inter-zonal power flows computed using various algorithms as described in~\cite{ward1949equivalent, dimo1975nodal, oh2009new, shi2012improved, shi2014novel}. Our proposed algorithm has superior performance with $0.2$~MW mean absolute error (MAE) in inter-zonal power flows relative to the true AC solution. In contrast, other methodologies have MAE values that are one to two orders-of-magnitude worse (ranging from $3.9$ MW to $37.7$ MW). Matching~\cite{ward1949equivalent, dimo1975nodal, oh2009new, shi2012improved, shi2014novel}, the resistance of the transmission lines has been set to zero for this analysis. The proposed algorithm's accuracy advantages would likely be more pronounced otherwise due to the capability to account for losses via the bias parameters $\boldsymbol{\gamma}_R$ and $\boldsymbol{\rho}_R$, a feature not shared by the other methods. Moreover, the injection scenario presented here serves as the foundational case upon which other methods base their calculations of transmission line reactances. In contrast, our algorithm has been trained across a range of injection scenarios, thereby enhancing its ability to accurately compute inter-zonal flows under diverse conditions. To verify this, we trained the proposed algorithm using only the base case scenario (i.e., the given power injections above), and the MAE was reduced to zero.
\begin{table}[t]
\centering
\caption{Optimized Values of \( \mathbf{b}_R \), \( \boldsymbol{\gamma}_R \), and \(\boldsymbol{\rho}_R \) for the 4-bus Equivalent}
\vspace{-1em}
\label{tab:parameters}
\begin{tabular}{lccc}
\hline
& \( \mathbf{b}^{opt}_R \)& \( \boldsymbol{\gamma}^{opt}_R \)& \(\boldsymbol{\rho}^{opt}_R \) \\
\hline
Bus at Zone II & - & 0.023264 & - \\
Bus at Zone III & - & -0.001685 & - \\
Bus at Zone IV & - & -0.021608 & - \\
\hline
Line \( Zone~I-Zone~II \) & 12.715 & - & 0.0205 \\
Line \( Zone~I-Zone~IV \) & 14.081 & - & -0.0215 \\
Line \( Zone~II-Zone~III \) & 8.733 & - & -0.0177 \\
Line \( Zone~II-Zone~IV \) & 6.870 & - & 0.0628 \\
Line \( Zone~III-Zone~IV \) & 7.597 & - & -0.0190 \\
\hline
\end{tabular}
\vspace{-1em}
\end{table}

\begin{table}[t]
\centering
\footnotesize
\caption{Inter-zonal Power Flows (MW) Using Different Algorithms}
\vspace{-1em}
\setlength{\tabcolsep}{4pt} 
\renewcommand{\arraystretch}{1.1}
\label{tab:comparison}
\begin{tabularx}{\columnwidth}{ccccccc}
\hline
\textbf{Flow}& \textbf{Ward} & \textbf{REI} & \textbf{Ref.} & \textbf{Refs.} &\textbf{Proposed} & \textbf{Actual AC}  \\ 
& \textbf{\cite{ward1949equivalent}} & \textbf{\cite{dimo1975nodal}} & \textbf{\cite{oh2009new}} & \textbf{\cite{shi2012improved, shi2014novel}} & \textbf{algorithm} & \textbf{solution} \\ \hline \hline
$p_{I->II}$    & -271.4 & -316.3 & -244.8 & -232.8 & -238.8 &-238.4   \\ \hline
$p_{I->IV}$    & -128.6 & -130.3 & -155.3 & -167.2 & -161.2 & -161.6  \\ \hline
$p_{II->III}$  & 10.7   & -42.8  & 1.8    & 5.7    & 2.5    & 2.6     \\ \hline
$p_{II->IV}$   & 17.9   & 26.4   & 53.5   & 61.5   & 58.8   & 58.9    \\ \hline
$p_{III->IV}$  & 60.7   & 53.9   & 51.8   & 55.7   & 52.5   & 52.6    \\ \hline
$p_{I->III}$   & N/A    & 46.6   & N/A    & N/A    & N/A    & N/A     \\ \hline
\textbf{MAE} & 24.6 & 37.7 & 3.9 & 4.0 &0.2& N/A \\ \hline
\end{tabularx}
\vspace{-1em}
\end{table}

\subsection{Scalability of the Proposed Algorithm}

To demonstrate the scalability of our proposed algorithm, we applied it to larger networks. Specifically, we tested it on the \texttt{IEEE 118-bus} system and a larger \texttt{4601-bus} system. In the \texttt{IEEE 118-bus} system, we divided the buses into $43$ zones, with a few buses designated as single-bus zones. Similarly, for the \texttt{4601-bus} system, we segmented the network into $805$ distinct zones. The detailed zone configurations for each system are provided in the appendix.

After optimizing the coefficient and bias parameters \( \mathbf{b}_R \), \( \boldsymbol{\gamma}_R \), and \(\boldsymbol{\rho}_R \) in the training phase,  these parameters are employed in the online phase on the test dataset (i.e., $2,000$ scenarios) to assess the loss function as defined in \eqref{eq:objective_function}.

\begin{figure*}[htbp]
\centering
\subfloat[\small MAE ]{
\includegraphics[width=0.485\textwidth]{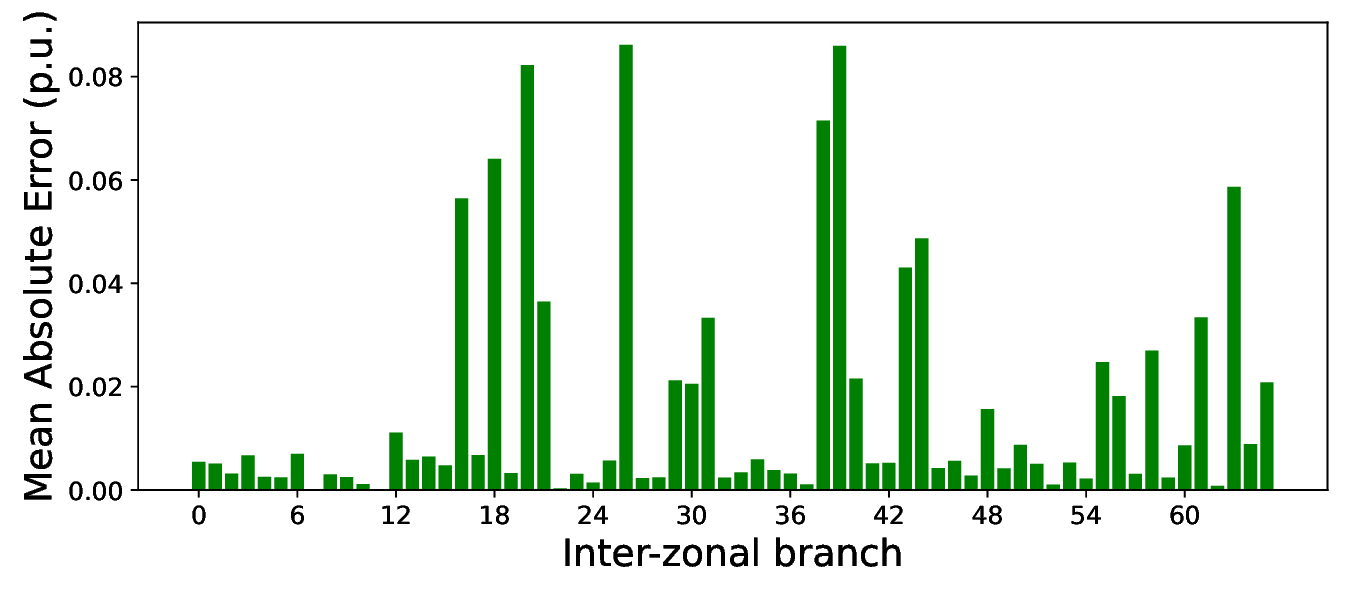}
\label{fig:Average_Power_Flow_Comparison}
}
\hfill
\subfloat[\small  Cumulative proportion]{
\includegraphics[width=0.455\textwidth]{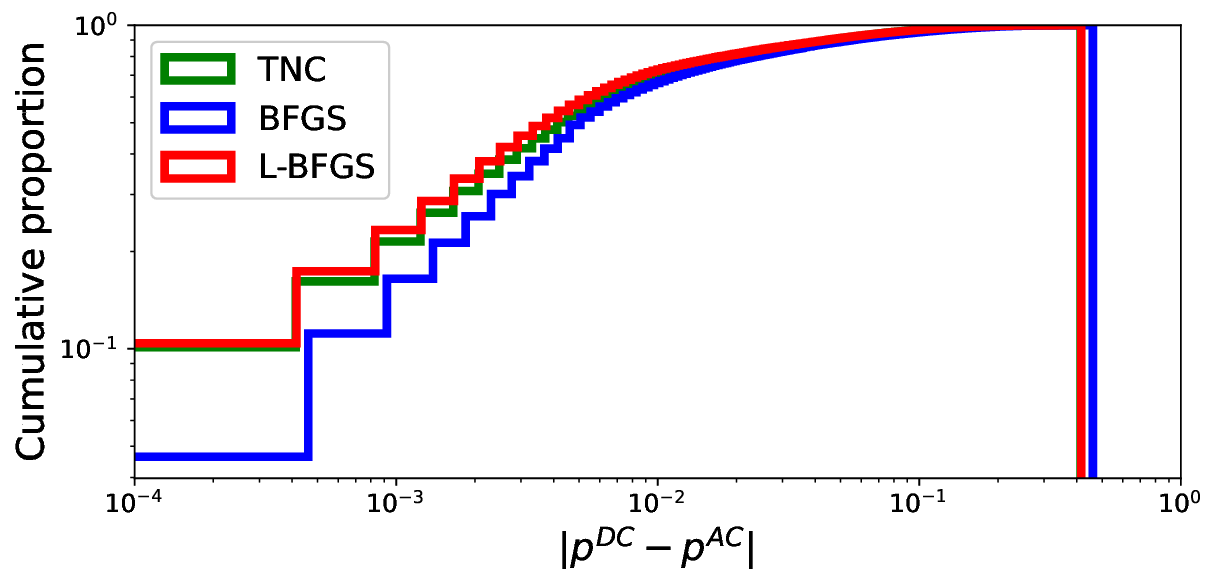}
\label{fig:118_Error_Power_Flow}
}
\vspace{-0.5em}
\caption{ (a) MAE between the AC and optimized DC power flow models for inter-zonal power flows in the reduced \texttt{IEEE 118-bus} system over $2,000$ test scenarios with $66$ inter-zonal branches. (b) Cumulative proportion of the absolute error between AC and DC power flows.}
\label{fig:Comparison}
\vspace{-1em}
\end{figure*}

As shown in Table~\ref{table:combined_loss}, the TNC method achieves the lowest loss for the \texttt{IEEE 118-bus} system, suggesting its efficacy in moderate-sized networks. Conversely, L-BFGS has the best results for the \texttt{4601-bus} system, suggesting its suitability for larger networks. The performance variation across different network sizes indicates that users should experiment with various optimization methods to find the most suitable one for their system. Different methods may yield better results depending on the specific network characteristics.
The training times for different methods suggest that while larger networks increase computational demands, the algorithm remains tractable. Overall, the algorithm's performance in minimizing loss values and maintaining practical training times underscores its potential for improving the computational speed of optimization problems and other applications of network equivalencing methods.

Fig.~\ref{fig:Average_Power_Flow_Comparison} showcases the MAE between AC and DC power flows, where the DC model utilizes optimized parameters \(\mathbf{b}^{opt}_R\), \(\boldsymbol{\gamma}^{opt}_R\), and \(\boldsymbol{\rho}^{opt}_R\). This analysis covers $66$ inter-zonal branches and $2,000$ scenarios in the \texttt{IEEE 118-bus} system, highlighting the high accuracy of the DC model with MAE values under $0.1$ per unit across all branches. Complementarily, Fig.~\ref{fig:118_Error_Power_Flow} illustrates the effectiveness of the optimization methods, demonstrating maximum errors below $0.413$, $0.416$, and $0.488$ per unit for TNC, L-BFGS, and BFGS methods, respectively, thereby emphasizing the performance advantages of our optimized parameters.

\begin{table}
\centering
\caption{Squared Two-Norm and $\infty$-Norm Loss Functions Using 2,000 test dataset with the L-BFGS, BFGS, and TNC Methods}
\vspace{-1em}
\setlength{\tabcolsep}{1.5pt} 
\renewcommand{\arraystretch}{1.1}
\resizebox{\columnwidth}{!}{%
\begin{tabular}{lcccccccccc}
\toprule
\textbf{Test case} & \multicolumn{3}{c}{\textbf{Squared Two-Norm Loss}} & \multicolumn{3}{c}{\textbf{$\infty$-Norm Loss}} & \multicolumn{3}{c}{\textbf{Training (s)}}\\ 
\cmidrule(lr){2-4} \cmidrule(lr){5-7} \cmidrule(lr){8-10}
  & L-BFGS & BFGS & TNC & L-BFGS &  BFGS  &  TNC & L-BFGS &  BFGS  &  TNC \\ 
\midrule
\midrule

 \texttt{6-bus}   &1.175  &1.420  &$\mathbf{1.098}$   &$\mathbf{0.114}$  &0.115  &0.117   &5 &$\mathbf{3}$  &4   \\
 \texttt{118-bus} &2.420 &3.538  &$\mathbf{2.392}$    &0.416 &0.488  &$\mathbf{0.413}$   &131  &$\mathbf{33}$  &45   \\
 \texttt{4601-bus}&$\mathbf{2.758}$  &2.791  &2808.045   &$\mathbf{1.485}$  &1.517  &18.454  &$\mathbf{346}$  &2190  &1306  \\
\bottomrule
\end{tabular}
}
\begin{tablenotes}
 \item[*] \scriptsize \hspace*{-2em} The best performing method (smallest loss function, fastest) is bolded for each test case.
\end{tablenotes}
\label{table:combined_loss}
\vspace{-1em}
\end{table}

\section{Conclusion}
\label{sec:Conclusion}

This paper presents an algorithm designed to optimize the parameters of a reduced network, with the aim of tailoring the DC power flow model to closely align with the results of the AC power flow equations on the original network. Unlike conventional methods that focus on a single nominal operating point, our algorithm considers a range of operations, enhancing the versatility and accuracy of the DC power flow model in varied conditions. The algorithm adjusts coefficient and bias parameters to minimize the discrepancies between DC and AC power flow solutions across scenarios. We use optimization methods such as BFGS, L-BFGS, and TNC to fine-tune these parameters offline, thereby improving the precision of online DC power flow approximations for various optimization applications. Our numerical results demonstrate substantial improvements in accuracy across multiple test cases.

Future work will explore refining network reduction by using a predefined number of buses and lines per zone instead of a single bus. This modification seeks to more effectively accommodate distributed optimization settings, potentially facilitating the safe sharing of simplified models between adjacent areas without revealing the complete model of each area.

\bibliographystyle{IEEEtran}
\bibliography{refs}

\appendix[Defined Zones for Larger Networks]
\label{sec:appendix}
For the \texttt{IEEE 118-bus} system, the zones are: 
\( \{1, \ldots, 32, 113, 114, 115, 117\} \), 
\( \{33, \ldots, 37, 39, \ldots, 42\} \), 
\( \{43, \ldots, 49\} \), 
\( \{50, \ldots, 61, 63, 64\} \), 
\( \{62, 65, \ldots, 69, 116\} \), and 
\( \{24, 38, 70, \ldots, 75, 118\} \). 
Any bus not included in these groupings is considered its own single-bus zone.
For the \texttt{4601-bus} system, the zones are: \( \{500, \ldots, 999\} \), \( \{1000, \ldots, 1999\} \),\(\{2000, \ldots, 2999\} \), \( \{3000, \ldots, 3999\} \), and \( \{4300, \ldots, 4601\} \).

\end{document}